\documentclass[aps,prx,twocolumn,superscriptaddress,showkeys,preprintnumbers,amsmath,amssymb]{revtex4-1}  
\usepackage{graphicx}  
\usepackage{dcolumn}   
\usepackage{bm}        
\usepackage[latin1]{inputenc}
\usepackage{color}
\usepackage{eqnarray}

\begin{document}


\title{Bulk Fermi-surface of the Weyl type-II semi-metal candidate $\gamma$-MoTe$_2$}
\author{D. Rhodes}
\affiliation{National High Magnetic Field Laboratory, Florida State University, Tallahassee-FL 32310, USA}
\affiliation{Department of Physics, Florida State University, Tallahassee-FL 32306, USA}
\author{R. Sch\"{o}nemann}
\affiliation{National High Magnetic Field Laboratory, Florida State University, Tallahassee-FL 32310, USA}
\author{N. Aryal}
\affiliation{National High Magnetic Field Laboratory, Florida State University, Tallahassee-FL 32310, USA}
\affiliation{Department of Physics, Florida State University, Tallahassee-FL 32306, USA}
\author{Q. Zhou}
\affiliation{National High Magnetic Field Laboratory, Florida State University, Tallahassee-FL 32310, USA}
\affiliation{Department of Physics, Florida State University, Tallahassee-FL 32306, USA}
\author{Q. R. Zhang}
\affiliation{National High Magnetic Field Laboratory, Florida State University, Tallahassee-FL 32310, USA}
\affiliation{Department of Physics, Florida State University, Tallahassee-FL 32306, USA}
\author{E. Kampert}
\affiliation{Dresden High Magnetic Field Laboratory (HLD-EMFL), Helmholtz-Zentrum Dresden-Rossendorf, 01328 Dresden, Germany}
\author{Y.-C. Chiu}
\affiliation{National High Magnetic Field Laboratory, Florida State University, Tallahassee-FL 32310, USA}
\affiliation{Department of Physics, Florida State University, Tallahassee-FL 32306, USA}
\author{Y. Lai}
\affiliation{National High Magnetic Field Laboratory, Florida State University, Tallahassee-FL 32310, USA}
\affiliation{Department of Physics, Florida State University, Tallahassee-FL 32306, USA}
\author{Y. Shimura}
\affiliation{National High Magnetic Field Laboratory, Florida State University, Tallahassee-FL 32310, USA}
\affiliation{University of Tokyo, Institute of Solid State Physics, Kashiwa, Chiba 2778581, Japan}
\author{G. T. McCandless}
\affiliation{The University of Texas at Dallas, Department of Chemistry and Biochemistry, Richardson, TX 75080 USA}
\author{J. Y. Chan}
\affiliation{The University of Texas at Dallas, Department of Chemistry and Biochemistry, Richardson, TX 75080 USA}
\author{D. W. Paley}
\affiliation{Department of Chemistry, Columbia University, New York, NY 10027 USA}
\affiliation{Columbia Nano Initiative, Columbia University, New York, NY 10027 USA}
\author{J. Lee}
\affiliation{CHESS, Cornell University, Ithaca, New York 14853, USA}
\author{A. D. Finke}
\affiliation{CHESS, Cornell University, Ithaca, New York 14853, USA}
\author{J. P. C. Ruff}
\affiliation{CHESS, Cornell University, Ithaca, New York 14853, USA}
\author{S. Das}
\affiliation{National High Magnetic Field Laboratory, Florida State University, Tallahassee-FL 32310, USA}
\affiliation{Department of Physics, Florida State University, Tallahassee-FL 32306, USA}
\author{E. Manousakis}
\affiliation{Department of Physics, Florida State University, Tallahassee-FL 32306, USA}
\affiliation{National High Magnetic Field Laboratory, Florida State University, Tallahassee-FL 32310, USA}
\author{L. Balicas}
\email[]{balicas@magnet.fsu.edu}
\affiliation{National High Magnetic Field Laboratory, Florida State University, Tallahassee-FL 32310, USA}
\affiliation{Department of Physics, Florida State University, Tallahassee-FL 32306, USA}
\date{\today}

\begin{abstract}
The electronic structure of semi-metallic transition-metal dichalcogenides, such as WTe$_2$ and orthorhombic $\gamma-$MoTe$_2$, are claimed to contain pairs of Weyl points or linearly touching electron and hole pockets associated with a non-trivial Chern number. For this reason, these compounds were recently claimed to conform to a new class, deemed type-II, of Weyl semi-metallic systems. A series of angle resolved photoemission experiments (ARPES) claim a broad agreement with these predictions detecting, for example, topological Fermi arcs at the surface of these crystals. We synthesized single-crystals of semi-metallic MoTe$_2$ through a Te flux method to validate these predictions through measurements of its bulk Fermi surface (FS) \emph{via} quantum oscillatory phenomena. We find that the superconducting transition temperature of $\gamma-$MoTe$_2$ depends on disorder as quantified by the ratio between the room- and low-temperature resistivities, suggesting the possibility of an unconventional superconducting pairing symmetry. Similarly to WTe$_2$, the magnetoresistivity of $\gamma-$MoTe$_2$ does not saturate at high magnetic fields and can easily surpass $10^{6}$ \%. Remarkably, the analysis of the de Haas-van Alphen (dHvA) signal superimposed onto the magnetic torque, indicates that the geometry of its FS is markedly distinct from the calculated one. The dHvA signal also reveals that the FS is affected by the Zeeman-effect precluding the extraction of the Berry-phase. A direct comparison between the previous ARPES studies and density-functional-theory (DFT) calculations reveals a disagreement in the position of the valence bands relative to the Fermi level $\varepsilon_F$. Here, we show that a shift of the DFT valence bands relative to $\varepsilon_F$, in order to match the ARPES observations, and of the DFT electron bands to explain some of the observed dHvA frequencies, leads to a good agreement between the calculations and the angular dependence of the FS cross-sectional areas observed experimentally. However, this relative displacement between electron- and hole-bands eliminates their crossings and, therefore, the Weyl type-II points predicted for $\gamma-$MoTe$_2$.
\end{abstract}

\maketitle

\section{Introduction}

The electronic structure of the transition-metal dichalcogenides (TMDs) belonging to the orthorhombic and non-centrosymmetric $Pmn2_1$ space group, e.g. WTe$_2$, were recently recognized as candidates for possible topologically non-trivial electronic states. For instance, their  monolayer electronic bands were proposed to be characterized by a non-trivial $Z_2 = 1$ topological invariant based on the parity of their valence bands, making their monolayers good candidates for a quantum spin Hall insulating ground-state \cite{TP_transition}. This state is characterized by helical edge states that are protected by time-reversal symmetry from both localization and elastic backscattering. Hence, these compounds could provide a platform for realizing low dissipation quantum electronics and spintronics \cite{TP_transition, MacDonald}.

However, the majority of gapped TMDs, such as semiconducting MoS$_2$ or WSe$_2$, crystallize either in a trigonal prismatic coordination or in a triclinic structure with octahedral coordination \cite{review1,review2} as is the case of ReS$_2$. Those crystallizing in the aforementioned orthorhombic phase, e.g. WTe$_2$, are semi-metals albeit displaying remarkable transport properties such as an enormous, non-saturating magnetoresistivity \cite{cava}. Strain is predicted to open a band gap \cite{TP_transition} in WTe$_2$, which might make it suitable for device development. In fact, simple exfoliation of its isostructural $\gamma-$MoTe$_2$ compound (where $\gamma$ refers to the orthorhombic semi-metallic phase) into thin atomic layers was claimed to induce a band gap \cite{MoTe2_MI_transition} in the absence of strain. Such a transition would contrast with band structure calculations finding that WTe$_2$ should remain semi-metallic when exfoliated down to a single atomic layer \cite{Lv}. The insulating behavior reported for a few atomic layers of WTe$_2$ was ascribed to an increase in disorder due to its chemical instability in the presence of humidity which would induce Anderson localization \cite{Morpurgo}, although more recently it was claimed to be intrinsic from transport measurements on encapsulated few-layered samples \cite{Cobden}.

Orthorhombic $\gamma-$MoTe$_2$ and its isostructural compound WTe$_2$ were also claimed, based on density functional theory calculations, to belong to a new class of Weyl semi-metals, called type-II, which is characterized by a linear touching between hole and electron Fermi surface pockets \cite{bernevig,felser,bernevig2, Hasan}. As for conventional Weyl points \cite{Weyl1, Weyl2}, these Weyl type-II points would also act as topological charges associated with singularities, i.e., sources and sinks, of Berry-phase pseudospin \cite{bernevig,felser,bernevig2,Hasan} which could lead to anomalous transport properties. A series of recent angle-resolved photoemission spectroscopy (ARPES) measurements \cite{ARPES_Huang, ARPES_Deng, ARPES_Jiang, ARPES_Liang, ARPES_Xu, ARPES_Tamai, ARPES_Belopolski, thirupathaiah} claim to observe a good overall agreement with these predictions. These studies observe the band crossings predicted to produce the Weyl type-II points, which would be located slightly above the Fermi-level, as well as the Fermi arcs projected on the surface of this compound \cite{ARPES_Huang, ARPES_Deng, ARPES_Jiang, ARPES_Liang, ARPES_Xu, ARPES_Tamai, ARPES_Belopolski, thirupathaiah}.

Here, motivated by the scientific relevance and the possible technological implications of the aforementioned theoretical predictions \cite{bernevig,felser,bernevig2,Hasan, TP_transition}, we evaluate, through electrical transport and torque magnetometry in bulk single-crystals, the electronic structure at the Fermi level and the topological character of orthorhombic $\gamma-$MoTe$_2$. Our goal is to contrast our experimental observations with the theoretical predictions and the reported ARPES results in order to validate their findings. This information could, for example, help us predict the electronic properties of heterostructures fabricated from single- or a few atomic layers of this compound. An agreement between the calculated geometry of the FS of $\gamma-$MoTe$_2$ with the one extracted from quantum oscillatory phenomena, would unambiguously support the existence of Weyl nodes in the bulk \cite{Weyl1,Weyl2} and, therefore, the existence of related non-trivial topological surface states or Fermi arcs \cite{Weyl2, Hasan, bernevig,felser,bernevig2,Hasan}. However, quantum oscillatory phenomena from $\gamma-$MoTe$_2$ single-crystals reveals a Fermi surface whose geometry is quite distinct from the one predicted by the DFT calculations based on its low temperature crystallographic structure. The extracted Berry-phase is found to be field-dependent. Still one does not obtain evidence for the topological character predicted for this compound when the Berry-phase is evaluated at low fields. Here, we show that shifts in the relative position of the electron and hole bands, implied by previous ARPES studies \cite{ARPES_Huang, ARPES_Deng, ARPES_Jiang, ARPES_Liang, ARPES_Xu, ARPES_Tamai, ARPES_Belopolski, thirupathaiah}, can replicate the angular dependence of the observed Fermi surface cross-sectional areas. However, these band shifts imply that the valence and electron bands would no longer cross and, therefore, that $\gamma-$MoTe$_2$ would not display the predicted Weyl type-II semi-metallic state.
\begin{figure*}[htp]
\begin{center}
    \includegraphics[width=18 cm]{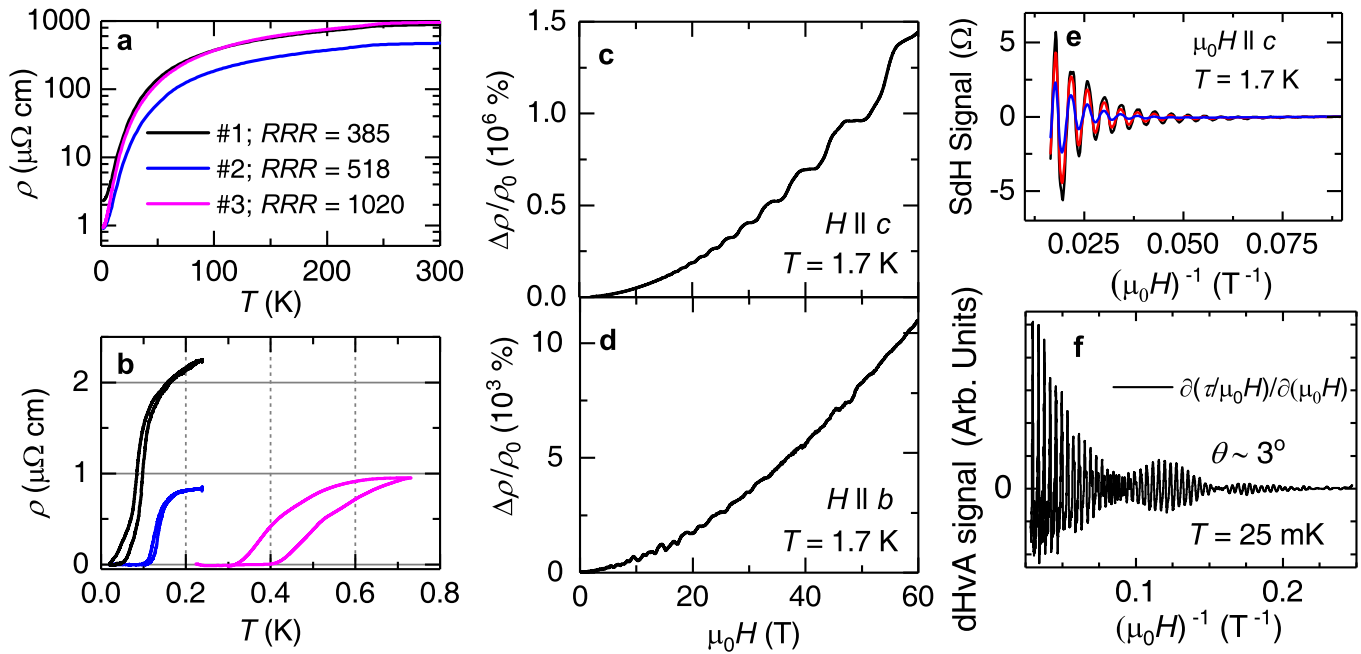}
    \caption{(a) Resistivity $\rho$, for currents flowing along the $a-$axis, as a function of the temperature $T$ for three representative single crystals displaying resistivity ratios $\rho(300 \text{ K})/\rho(2) \text{ K}$ between 380 and $\gtrsim 1000$. (b) $\rho$ as a function of $T$ for each single-crystal indicating that $T_c$ depends on sample quality: it increases from an average transition middle point value of $\sim 130$ mK for the sample displaying the lowest ratio to $\sim 435$ mK for the sample displaying the highest one. The apparent hysteresis is due to a non-ideal thermal coupling between the single-crystals, the heater and the thermometer. (c) $\rho$ as a function of the field $H$ applied along the \emph{c}-axis at a temperature $T = 1.7$ K for a fourth crystal characterized by a resistivity ratio of $\sim 207$. Notice i) the non-saturation of $\rho(H)$ and ii) that $\Delta \rho (\mu_0 H)/ \rho_0 = (\rho(\mu_0 H)- \rho_0)/ \rho_0$, where $\rho_0 = \rho(\mu_0 H=0 \text{ T}, T = 2 \text{ K})$  surpasses $1.4 \times 10^6$ \% at $\mu_0 H = 60$ T. (d) $\rho$ as a function of $\mu_0 H$ applied along the $b-$axis also at $T=1.7$ K and for the same single-crystal. (e) Shubnikov-de Haas signal superimposed onto the magnetoresistivity for $\mu_0 H \| c-$axis and for three temperatures, $T = 8$ K (blue line), 4.2 K (red line) and 1.7 K (black line), respectively. (f) Oscillatory signal (black line) superimposed onto the magnetic susceptibility $\Delta \chi = \partial (\tau/\mu_0 H)/\partial \mu_0 H$, where $\tau$ is the magnetic torque.}
\end{center}
\end{figure*}
\begin{figure*}[ht]
\begin{center}
    \includegraphics[width=14cm]{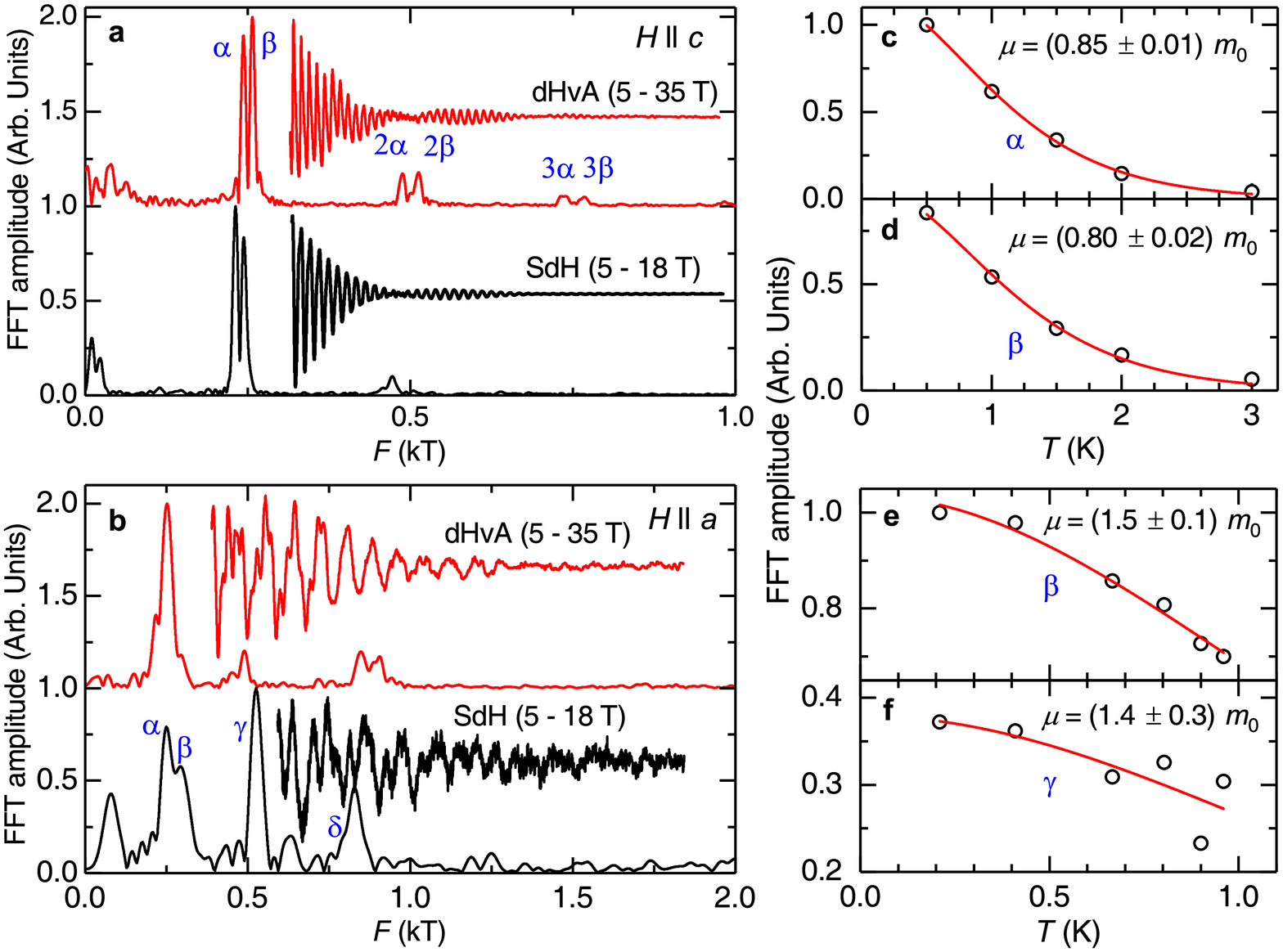}
    \caption{(a) Typical de Haas-van Alphen (red trace) and Shubnikov-de Haas (black trace) signals superimposed onto the magnetic torque and the magnetoresistivity respectively, for fields aligned along the \emph{c}-axis of $\gamma-$MoTe$_2$ single-crystals at $T \simeq 30$ mK. The same panel shows the FFT spectra for each signal revealing just two main frequencies or FS cross-sectional areas. (b) dHvA and SdH signals and corresponding FFT spectra obtained from the same single-crystals for $H$ aligned nearly along the \emph{a}-axis of each single-crystal. (c) and (d), Amplitude of the peaks observed in the FFT spectra for $H\|c$-axis and as a function of $T$ including the corresponding fits to the LK formula from which the effective masses are extracted. (e) and (f) Amplitude of two representative peaks observed in the FFT spectra for $H\|a$-axis and as a function of $T$ with the corresponding fits to the LK formula to extract their effective masses.}
\end{center}
\end{figure*}

\section{Methods and Experimental Results}

Very high quality single crystals of monoclinic $\beta-$MoTe$_2$ were synthesized through a Te flux method: Mo, 99.9999\%, and Te 99.9999 \% powders were placed in a quartz ampoule in a ratio of 1:25 heated up to 1050 $^{\circ}$C and held for 1 day. Then, the ampoule was slowly cooled down to 900 $^{\circ}$C and centrifuged. The ``as harvested" single-crystals were subsequently annealed for a few days at a temperature gradient to remove the excess Te. Magneto-transport measurements as a function of temperature were performed in a Physical Property Measurement System using a standard four-terminal configuration. Measurements of the Shubnikov-de Haas (SdH) and the de Haas-van Alphen (dHvA) effects were performed in dilution refrigerator coupled to a resistive Bitter magnet, with the samples immersed in the $^3$He-$^4$He mixture. Measurements of the dHvA-effect were performed \emph{via} a torque magnetometry technique, i.e. by measuring the deflection of a Cu-Be cantilever capacitively. Electrical transport measurements in pulsed magnetic fields were performed at the Dresden High Magnetic Field Laboratory using a 62 T magnet with a pulse duration of 150 ms. The sample temperature was controlled using a $^4$He bath cryostat (sample in He atmosphere) with an additional local heater for temperatures above 4.2 K. Synchrotron based X-ray measurements were performed in three single-crystals at the CHESS-A2 beam line using a combination of photon energies and cryogenic set-ups. The crystallographic data was reduced with XDS \cite{XDS}. The structures were solved with direct methods using SHELXS \cite{shelx}. Outlier rejection and absorption correction was done with SADABS. Least squares refinement on the intensities were performed with SHELXL \cite{shelx}. For additional detailed information on the experimental set-ups used, see Supplemental Information file \cite{supplemental}.

As illustrated by Fig. 1(a), the as synthesized single-crystals display resistivity ratios \emph{RRR} = $\rho(T = 300 \text{ K})/\rho(T = 2 \text{ K})$ ranging from $380$ to $> 2000$ which is one to two orders of magnitude higher than the $RRR$ values currently in the literature (see, for example, Ref. \onlinecite{MoTe2_SC}). Although not clearly visible in Fig. 1(a) due to its logarithmic scale, a hysteretic anomaly is observed in the resistivity around $ 240 $ K corresponding to the monoclinic to orthorhombic structural transition which stabilizes what we denominate as the orthorhombic $\gamma-$MoTe$_2$ phase. For a clearer exposure of this transition and related hysteresis, see Ref. \onlinecite{qiong}. These  single-crystals were subsequently measured at much lower temperatures allowing us to determine their superconducting transition temperature $T_c$. Remarkably, and as seen in Fig. 1(b), we find that $T_c$ depends on sample quality, increasing considerably as the $RRR$ increases, suggesting that structural disorder suppresses $T_c$. For these measurements, particular care was taken to suppress the remnant field of the superconducting magnet since the upper critical fields are rather small (see Supplemental Fig. S1 \cite{supplemental}). The sample displaying the highest $RRR$ and concomitant $T_c$ was measured in absence of a remnant field. To verify that these differences in $T_c$ are not due to a poor thermal coupling between the sample and the thermometers, $T_c$ was measured twice by increasing and decreasing $T$ very slowly. The observed hysteresis is small relative to $T_c$ indicating that the measured $T_c$s are not an artifact. The values of the residual resistivities $\rho_0$ depend on a careful determination of the geometrical factors such as the size of the electrical contacts. Therefore, the $RRR$ provides a more accurate determination of the single-crystalline quality. In the past, the suppression of $T_c$ by impurities and structural defects was systematically taken as evidence for unconventional superconductivity \cite{andy, satoru1, satoru2}, e.g. triplet superconductivity \cite{Maeno} in Sr$_2$RuO$_4$. Nevertheless, the fittings of the upper-critical fields $H_{c2}$ to a conventional Ginzburg-Landau expression, shown in Fig. S1 \cite{supplemental}, points towards singlet pairing.

We have also evaluated the quality of our single crystals through Hall-effect \cite{qiong} and heat capacity measurements (see, Supplemental Fig. S2 \cite{supplemental}). Hall-effect reveals a sudden increase in the density of holes below $T = 40$ K, suggesting a possible temperature-induced Lifshitz-transition. While the heat capacity reveals a broad anomaly around $T = 66$ K, well-below its Debye temperature ($\Theta_D \simeq 120$ K), that would suggest that the structural degrees of freedom continue to evolve upon cooling below $T = 100$ K. Given that such structural evolution could affect the electronic band structure predicted for $\gamma-$MoTe$_2$ \cite{felser,bernevig2,Hasan}, we performed synchrotron X-ray scattering down to $ \sim 12$ K (see, Supplemental Fig. S3 \cite{supplemental}). We observe some variability in the lattice constants extracted among several single-crystals and a sizeable hysteresis in the range $125 \text{K} \leq T \leq 250 \text{K}$ associated with the structural transition observed at $T \simeq 250$ K, but no significant evolution in the crystallographic structure below 100 K. As we discuss below, there are negligible differences between the electronic bands calculated with the crystal structures collected at 100 K and at 12 K, respectively.

Figures 1(c) and 1(d) display the change in magnetoresistivity $\Delta \rho (\mu_0 H)/\rho_0 = (\rho(\mu_0 H)-\rho_0)/\rho_0$ as a function of the field $\mu_0 H$ for a crystal characterized by $RRR \sim 450$ when the electrical current flows along the crystalline \emph{a}-axis and the field is applied either along the \emph{c}- or the \emph{b}-axes, respectively. Similarly to WTe$_2$, for both orientations $\Delta \rho /\rho_0$ shows no sign of saturation under fields all the way up to 60 T while surpassing $1 \times 10^{6}$ \% for $\mu_0 H \| c$-axis \cite{cava}. For WTe$_2$ such anomalous magnetoresistivity was attributed to compensation between the density of electrons and holes \cite{cava,pletikosic,pippard}. Nevertheless, there are a number of subsequent observations \cite{Daniel} contradicting this simple scenario, such as i) a non-linear Hall response \cite{Joe}, ii) the suppression of the magnetoresistivity at a pressure where the Hall response vanishes \cite{WTe2_SC_1} (i.e. at perfect compensation), and iii) the observation of a pronounced magnetoresistivity in electrolyte gated samples with a considerably higher density of electrons with respect to that of holes \cite{Fuhrer}.  It remains unclear if the proposed unconventional electronic structure \cite{bernevig,felser,bernevig2} would play a role on the giant magnetoresistivity of WTe$_2$, while its measured FS differs from the calculated one \cite{Daniel,behnia}. In contrast, we have previously shown that $\gamma$-MoTe$_2$ indeed is a well compensated semi-metal \cite{qiong}. The slightly smaller
magnetoresistivity of $\gamma-$MoTe$_2$ relative to WTe$_2$ is attributable to heavier effective effective masses, according to de Haas-van Alphen-effect discussed below, or concomitantly lower mobilities.

The best $\gamma-$MoTe$_2$ samples, i.e. those with $RRR \geq 2000$, display even more pronounced $\Delta \rho/\rho_0$ under just $\mu_0 H \simeq 10$ T. The oscillatory component superimposed on the magnetoresistivity corresponds to the Shubnikov-de Haas (SdH) effect resulting from the Landau quantization of the electronic orbits. Figure 1(e) shows the oscillatory, or the SdH signal as a function of inverse field $(\mu_0H)^{-1}$ for three temperatures. The SdH signal was obtained by fitting the background signal to a polynomial and subtracting it. Notice how for this sample and for $\mu_0 H \|$ \emph{c}-axis, the SdH signal is dominated by a single frequency. However for all subsequent measurements performed under continuous fields (discussed below) one observes the presence of two main frequencies very close in value, each associated to an extremal cross-sectional area $A$ of the FS through the Onsager relation $F = A(\hbar / 2\pi e)$ where $\hbar$ is the Planck constant and $e$ is the electron charge. To illustrate this point, we show in Fig. 1(f) the oscillatory signal extracted from the magnetic torque, i.e. $\mathbf{\tau} = \mathbf{M} \times \mu_0 \mathbf{H}$, or the de Haas-van Alphen effect (dHvA) collected from a $\gamma-$MoTe$_2$ single-crystal for fields aligned nearly along its \emph{c}-axis. Here $M = \chi \mu_0 H$ is the magnetization and $\chi(\mu_0 H, T)$ is its magnetic susceptibility. Figure 1(f) shows the oscillatory component of the magnetic susceptibility $\Delta \chi = \partial (\tau/\mu_0H)/ \partial(\mu_0H)$. The envelope of the oscillatory signal displays the characteristic ``beating" pattern between two close frequencies. This becomes clearer in the fast Fourier transform of the oscillatory signal shown below.
According to the Lifshitz-Onsager quantization condition \cite{kim,nagaosa}, the oscillatory component superimposed onto the susceptibility is given by:
\begin{eqnarray*}
\Delta \chi[(B)^{-1}] \propto \frac{T}{B^{5/2}}\sum_{l=1}^{\infty} \frac{\exp^{-l \alpha \mu T_D/ B}\cos(l g \mu \pi/2)}{l^{3/2}\sinh(\alpha \mu T/B)}
\end{eqnarray*}
\begin{eqnarray}
\times \cos\left\{ 2\pi \left[ \left( \frac{F}{B}-\frac{1}{2}+\phi_B \right)l + \delta \right] \right\}
\end{eqnarray}
where $F$ is the dHvA frequency, $l$ is the harmonic index, $\omega_c$ the cyclotron frequency, $g$ the Land\'{e} \emph{g}-factor, $\mu$ the effective mass in units of the free electron mass $m_0$, and $\alpha$ is a constant. $\delta$ is a phase shift determined by the dimensionality of the FS which acquires a value of either $\delta = 0$ or $ \pm 1/8$ for two- and three-dimensional FSs \cite{kim,nagaosa,kopelevich}, respectively. $\phi_B$ is the Berry phase which, for Dirac and Weyl systems, is predicted to acquire a value $\phi_B = \pi$ \cite{kim,nagaosa,kopelevich}. Finally, $T_D= \hbar /(2 \pi k_B \tau $) is the so-called Dingle temperature from which one extracts $\tau$ or the characteristic quasiparticle scattering time.

In Supplementary Figs. S4 and S5 \cite{supplemental}, we discuss the extraction of the Berry-phase of $\gamma-$MoTe$_2$ via fits to Eq. (1) of the oscillatory signal shown in Fig. 1(f). However, the geometry of the FS of $\gamma-$MoTe$_2$ evolves slightly as the field increases due to the Zeeman-effect, which precludes the extraction of its Berry phase. More importantly, one cannot  consistently extract a value $\phi_B \simeq \pi$ when one limits the range in magnetic fields to smaller values in order to minimize the role of the Zeeman-effect. In other words, the dHvA-effect does not provide evidence for the topological character of $\gamma-$MoTe$_2$. Nevertheless, it does indicate that the Dingle temperature decreases as the field increases implying a field-induced increase in the quasiparticle lifetime. This effect should contribute to its large and non-saturating magnetoresistivity. We reported a similar effect for WTe$_2$ \cite{Daniel}.
\begin{figure*}[ht]
\begin{center}
    \includegraphics[width = 12 cm]{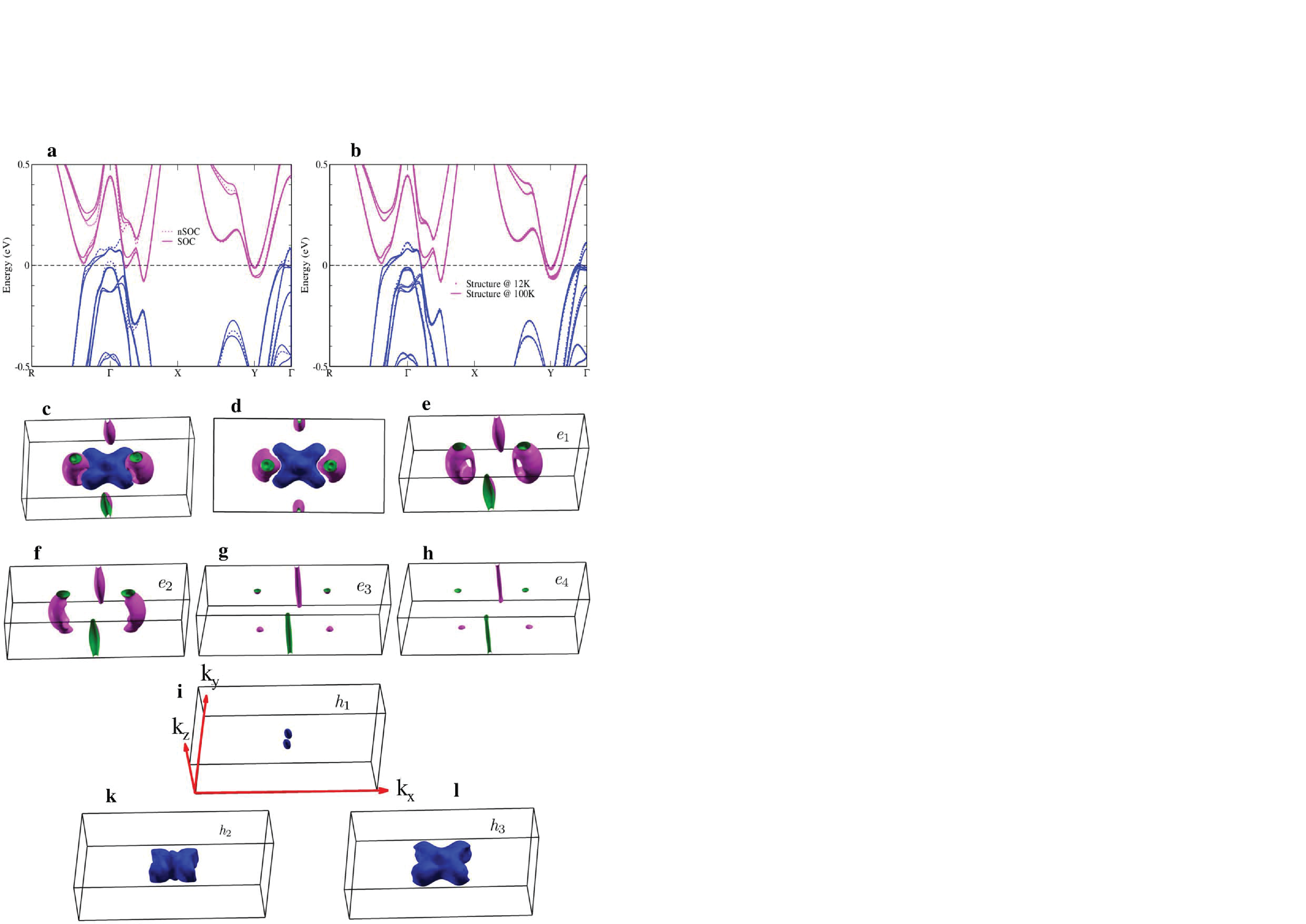}
    \caption{(a) Electronic band-structures of $\gamma-$MoTe$_2$ calculated with and without the inclusion of spin-orbit coupling (SOC). These calculations are based on the crystallographic structure measured at $T=100$ K. (b) Comparison between the calculated electronic bands based upon the crystallographic structures measured at $T = 100$ and $T = 12$ K, respectively. (c) and (d) Respectively, side and top views of the calculated FS. (e), (f), (g) and (h) Fermi surface sheets resulting from electron bands. Notice the marked two-dimensional character of several of the electron-like FS sheets. (i), (k) and (l), Hole-like sheets around the $\Gamma-$point.}
\end{center}
\end{figure*}
\begin{figure*}[htb]
\begin{center}
    \includegraphics[width = 10 cm]{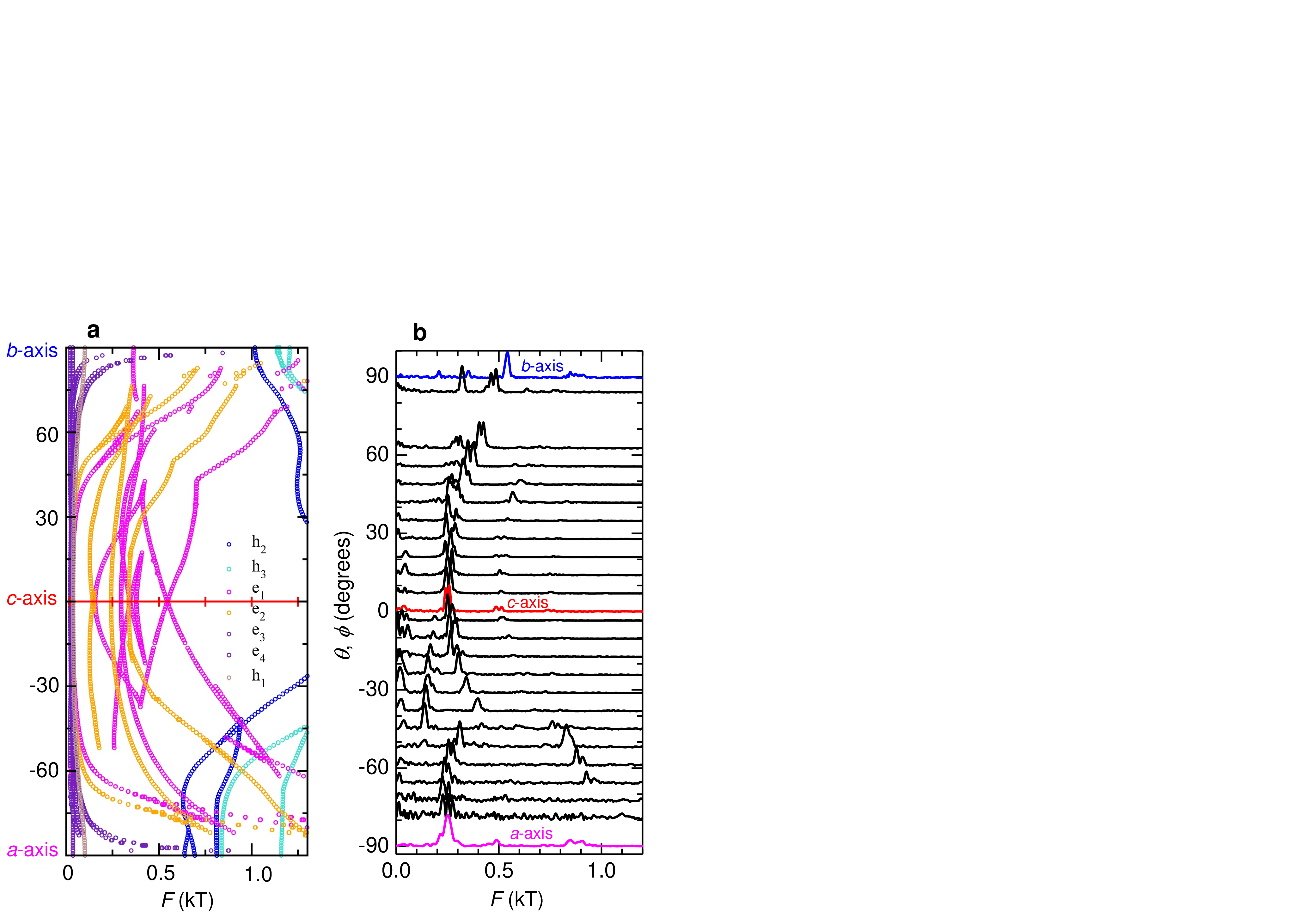}
    \caption{(a) Angular dependence of the FS cross-sectional areas or, through the Onsager relation, de Haas-van Alphen frequencies as with respect to the main crystallographic axes. (b) Experimentally observed dHvA spectra as a function of the frequency $F$ for several angles $\theta$ between the $c-$ and and the $b-$axis and for several values of the angle $-\phi$ between the $c-$ and the $a-$axes.}
\end{center}
\end{figure*}

Since the Berry-phase extracted from the dHvA-effect does not provide support for a topological semi-metallic state in $\gamma-$MoTe$_2$, it is pertinent to ask if the DFT calculations predict the correct electronic band-structure and related FS geometry for this compound, since both are the departing point for the predictions of Refs. \onlinecite{bernevig,felser,bernevig2}. To address this issue, we studied the dHvA-effect as a function of the orientation of the field with respect to the main crystallographic axes. Here, our goal is to compare the angular dependence of the cross-sectional areas determined experimentally, with those predicted by DFT.

Figures 2(a) and 2(b) display both the dHvA (red traces) and the SdH signals (black traces) measured in two distinct single crystals and for two field orientations, respectively along the $c-$ and the $a-$axes. As previously indicated, the dHvA and SdH signals were obtained after fitting a polynomial and subtracting it from the background magnetic torque and magnetoresistivity traces, respectively. The SdH signal was collected from a crystal displaying a $RRR \gtrsim 1000 $ at $T \simeq 25$ mK under fields up to 18 T, while the dHvA one was obtained from a crystal displaying $RRR \gtrsim 2000$ at $T \simeq 35$ mK under fields up to 35 T. Both panels also display the Fast Fourier transform (FFT) of the oscillatory signal. For fields along the $c-$axis, one observes two main peaks at $F_{\alpha} = 231$ T and at $F_{\beta} = 242$ T, as well as their first- and second harmonics and perhaps some rather small frequencies which could result from imperfect background subtraction. We obtain the same two dominant frequencies regardless of the interval in $H^{-1}$ used to extract the FFTs. Supplemental Fig. S6 \cite{supplemental} displays the dHvA signal for $H$ aligned nearly along the $b-$axis along with the corresponding FFT spectra which are again dominated by two prominent peaks. The observation of just two main frequencies for $\mu_0H \| c-$axis is rather surprising since, as we show below, DFT calculations, including the effect of the spin-orbit interaction, predict several pairs of electron-like corrugated cylindrical FSs along with pairs of smaller three-dimensional electron-like sheets in the First-Brillouin zone. Around the $\Gamma-$point, DFT predicts at least a pair of four-fold symmetric helix-like large hole sheets. This complex FS should lead to a rich oscillatory signal, contrary to what is observed. One might argue that the non-observation of all of the predicted FS sheets would be attributable to an experimental lack of sensitivity or to poor sample quality which would lead to low carrier mobility. Nevertheless, our analysis of the Hall-effect within a two-carrier model \cite{qiong}, yields electron- and hole-mobilities ranging between $10^4$ and $10^5$ cm$^2$/Vs at low $T$s which is consistent with both the small residual resistivities and the large resistivity ratios of our measured crystals. Given that the magnetic torque is particularly sensitive to the anisotropy of the FS, such high mobilities should have allowed us to detect most of the predicted FSs, particularly at the very low $T$s and very high fields used for our measurements. Hence, we conclude that the geometry of the FS ought to differ considerably from the one predicted by DFT.

In Figs. 2(c) and 2(d) we plot the amplitude of the main peaks observed in the FFT spectra for fields along the $c-$axis as a function of the temperature. Red lines are fits to the Lifshitz-Kosevich (LK) temperature damping factor, i.e. $x/\sinh x$ with $x = 14.69 \mu T/H$ and with $\mu$ being the effective mass in units of the free electron mass, from which we extract the masses associated with each frequency. As seen, for $H \| c-$axis one obtains $\mu_{\alpha} = 0.85$ $m_0$ and $\mu_{\beta}= 0.8$ $m_0$, which contrasts with the respective values obtained for $H \| a-$axis, namely $\mu_{\alpha, \beta} \simeq 1.5$ $m_0$, see Figs. 2(e) and 2(f). As previously mentioned for $\mu_0H \|b-$axis, we observe two main frequencies, but by reducing the $H^{-1}$ window to focus on the higher field region, we detect additional frequencies (See, Fig. S6 \cite{supplemental}) which are characterized by heavier effective masses, i.e. in the order of $2.5-2.9$ $m_0$. This indicates that $\gamma-$MoTe$_2$ displays a higher anisotropy in effective masses when compared to WTe$_2$ \cite{Daniel}, although these masses are consistent with its sizeable $\gamma_e$ coefficient. Supplemental Fig. S7 \cite{supplemental} displays several traces of the dHvA signal as functions of the inverse field for several angles between all three main crystallographic axes. These traces are used to plot the angular dependence of the FS cross-sectional areas in order to compare these with the DFT calculated ones.
\section{Comparison between experiments and the DFT calculations}
\begin{figure*}[htb]
\begin{center}
    \includegraphics[width = 15 cm]{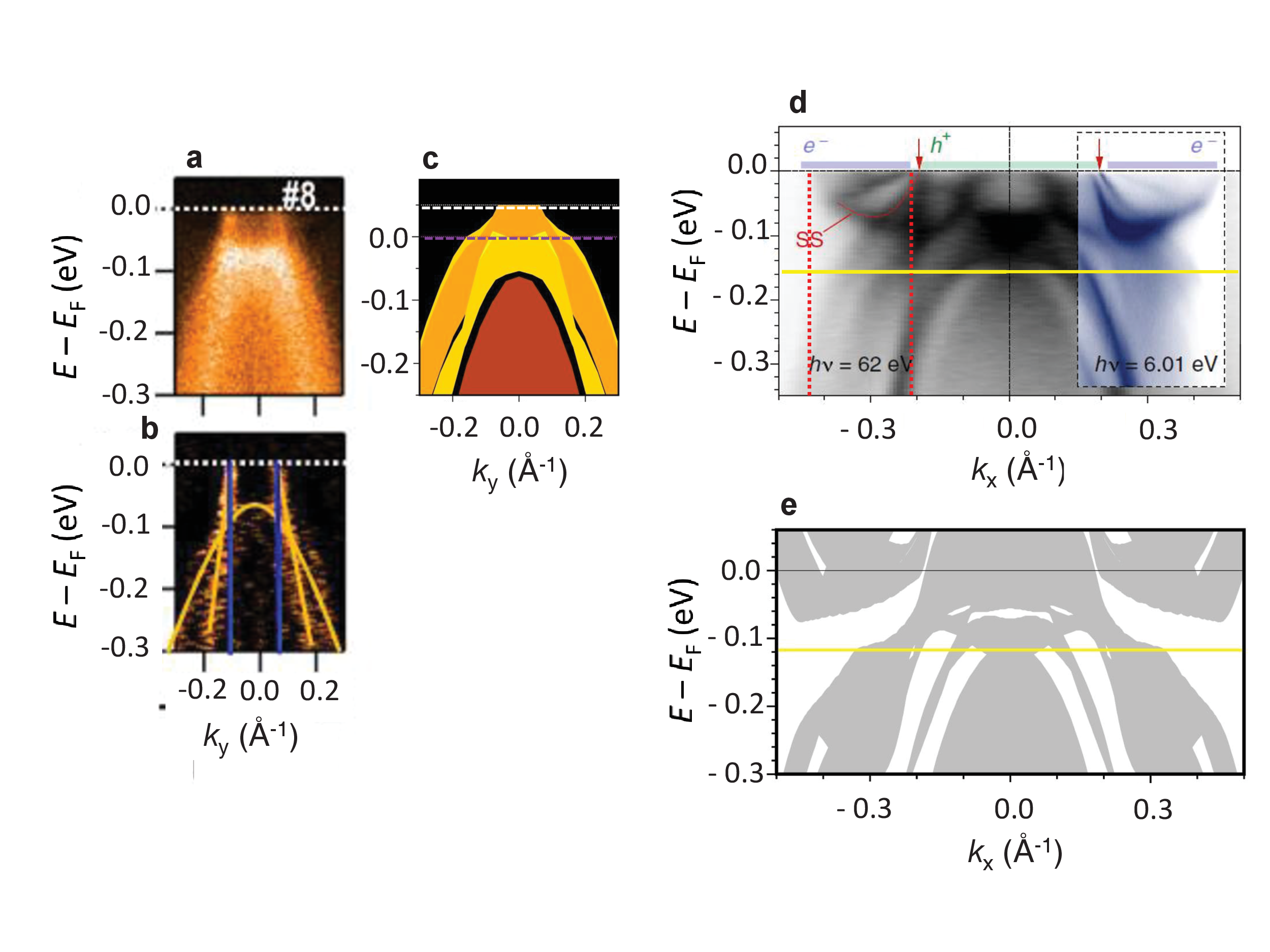}
    \caption{(a) ARPES energy distribution map along $k_y$ for $k_x=0$ from Ref. \onlinecite{thirupathaiah}. (b) Derivative of the energy distribution map. Vertical blue lines indicate the diameter of the observed hole-pocket, with its area corresponding to a frequency $F \sim 0.33$ kT.
    (c) Ribbons obtained from our DFT calculations showing the $k_z$-projection for all bulk bands which are plotted along the same direction as the ARPES energy distribution map in (a). Different colors are chosen to indicate distinct bands. Purple dotted line corresponds to the original position of the Fermi level $\epsilon_F$ according to DFT, while the white one corresponds to the position of $\epsilon_F$ according to ARPES. Notice that the ARPES bands are shifted by $\sim -50$ meV with respect to the DFT ones.  (d) ARPES energy distribution map corresponding to a cut along the $k_x-$direction with $k_y=0$, from Ref. \onlinecite{ARPES_Tamai}. Vertical red dotted lines indicate the diameter of the observed electron pockets or $\sim 0.2$ \AA$^{-1}$. SS stands for ``surface-state". (e) DFT Ribbons showing the $k_z$-projection for all bulk bands plotted along the same direction as the ARPES EDM in (d). In both panels black lines depict the original position of $\epsilon_F$ while the yellow lines are guides to the eyes illustrating the difference in energy between the top of the deepest DFT calculated hole-band and its equivalent according to ARPES. }
\end{center}
\end{figure*}
\begin{figure*}[htb]
\begin{center}
    \includegraphics[width=17 cm]{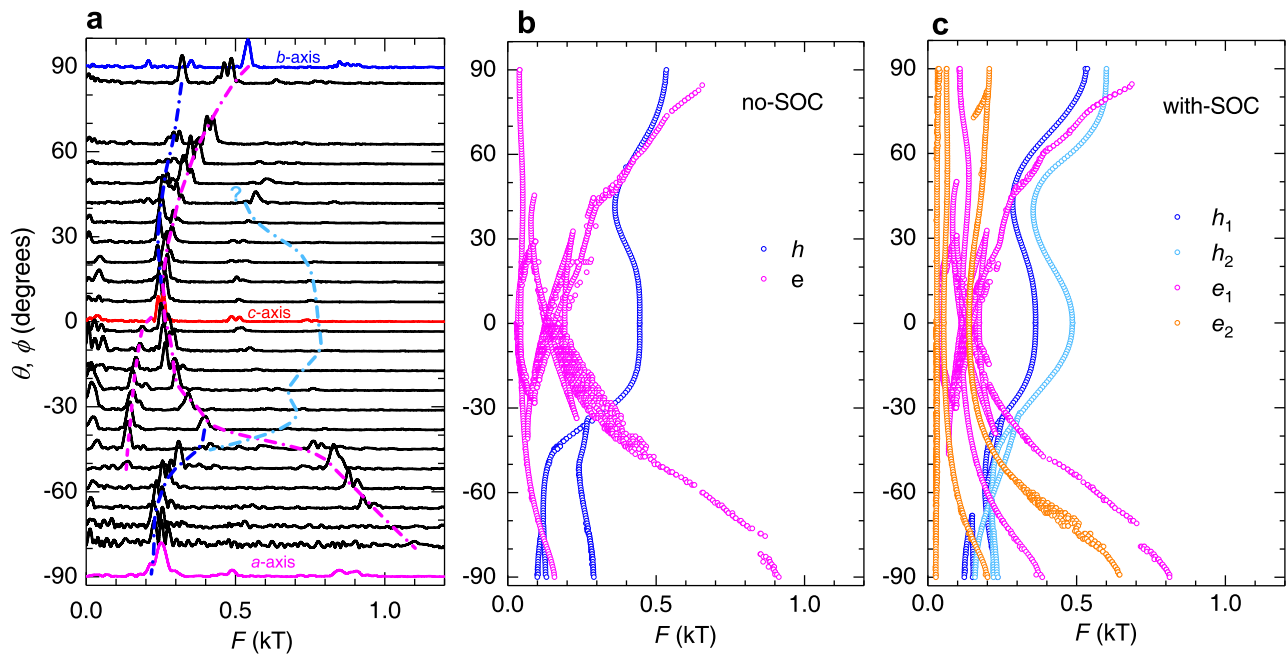}
\caption{(a) Experimentally observed dHvA spectra as a function of $F$ for several angles $\theta$ and $-\phi$, where the magenta and the blue lines act
as guides to the eyes and as identifiers of respectively, electron- and hole-like orbits according to the shifted band structure. Clear blue line depicts a possible hole-orbit associated with very small peaks in the FFT spectra. (b) Angular dependence of the dHvA orbits, or frequencies on the FS resulting from the shifted bands in absence of spin-orbit coupling, where magenta and blue markers depict electron- and hole-like orbits on the FS, respectively. Notice the qualitative and near quantitative agreement between the calculations and the experimental observations. (d) Angular dependence of the dHvA frequencies for the shifted electron and hole-bands in the presence of spin-orbit coupling. Here, electron-orbits are depicted by magenta and orange markers while the hole ones are indicated by blue and clear blue markers. The experimental data are better described by the non spin-orbit split bands.}
\end{center}
\end{figure*}
\begin{figure*}[htb]
\begin{center}
    \includegraphics[width = 14 cm]{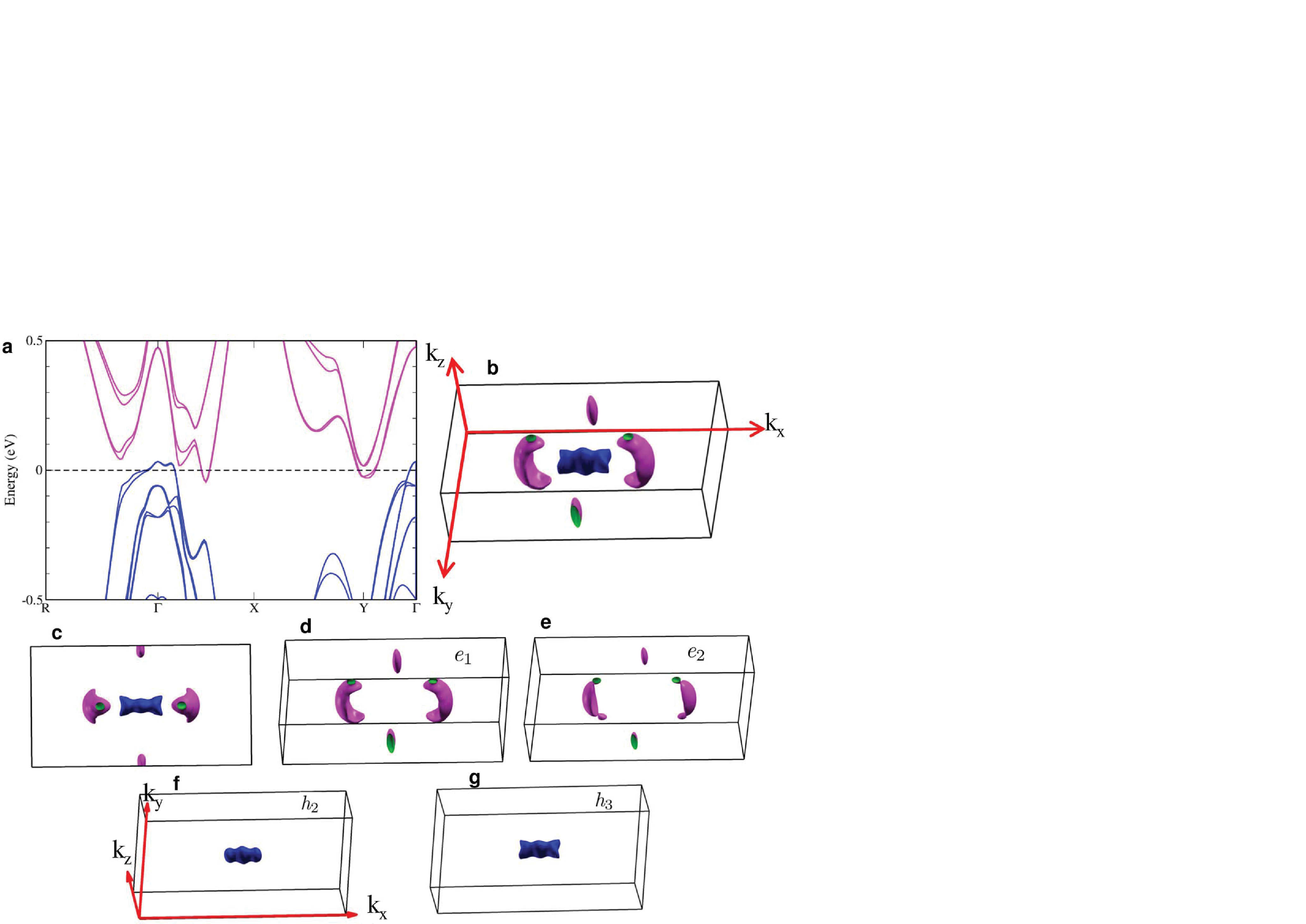}
    \caption{(a) Electronic band structure calculated with the inclusion of SOC after the hole-like bands have been shifted by $-50$ meV and the
electron ones by $+35$ meV with the goal of reproducing the observed dHvA frequencies and their angular dependence. Notice that these shifts suppress the crossings between the electron- and hole-bands and therefore the Weyl type-II points of the original band structure. (b) Fermi surface resulting from these band-shifts. (c) FS top view. (d) and (e) Electron-like FS sheets. (f) and (g) Hole-like sheets.  }
\end{center}
\end{figure*}

Several recent angle-resolved photoemission spectroscopy (ARPES) studies \cite{ARPES_Huang, ARPES_Deng, ARPES_Jiang, ARPES_Liang, ARPES_Xu, ARPES_Tamai,ARPES_Belopolski,thirupathaiah}
claim to find a broad agreement between the band structure calculations, the predicted geometry of the Fermi surface, the concomitant existence
of Weyl type-II points \cite{bernevig,felser,bernevig2}, and the related Fermi arcs on the surface states of $\gamma-$MoTe$_2$.
Several of these experimental and theoretical studies claim that the electronic structure of this compound is particularly sensitive to its precise crystallographic structure.
Inter-growth of the $2H-$phase or the temperature used to collect to X-ray diffraction data, typically around 100 to 230 K, are claimed to have a considerable effect on the calculations \cite{bernevig2, ARPES_Tamai}. Given the few frequencies observed by us, it is pertinent to ask if the mild evolution of the crystallographic structure as a function of the temperature shown in Fig. S3 \cite{supplemental} would affect the geometry of the FS of $\gamma-$MoTe$_2$. To address this question we performed a detailed angular-dependent study of the frequencies extracted from both the SdH and the dHvA effects in $\gamma-$MoTe$_2$ in order to compare these with the angular dependence of the FS cross-sectional areas predicted by the calculations.

In the subsequent discussion we compare the angular dependence of our dHvA frequencies with calculations performed with the Quantum Espresso \cite{QE} implementation of the density functional theory in the GGA framework including spin-orbit coupling (SOC). The Perdew-Burke-Ernzerhof (PBE) exchange correlation functional \cite{PBE} was used with fully relativistic norm-conserving pseudopotentials generated using the optimized norm-conserving Vanderbilt pseudopotentials as described in Ref. ~\onlinecite{ONCVPP}.
The 4\emph{s}, 4\emph{p}, 4\emph{d} and 5\emph{s} electrons of Mo and the 4\emph{d}, 5\emph{s} and  5\emph{p} electrons of Te were treated as valence electrons.
After careful convergence tests, the plane-wave energy cutoff  was taken to be 50 Ry and a $k-$point mesh of $20\times 12\times 6$ was used to sample the
reducible Brillouin Zone (BZ) used for the self-consistent calculation. The Fermi surfaces  were  generated  using  a more refined $k-$point mesh
of $45\times 25\times 14$. FS sheets were visualized  using the  XCrysden software \cite{xcrysden}. The related angular dependence of the quantum oscillation frequencies was calculated using the skeaf code \cite{skeaf}. As shown in Fig. 3 the results are very close to those obtained by using the VASP and the Wien2K implementations of DFT (see, Fig. S8 in SI \cite{supplemental}), and also to those reported by Refs. \onlinecite{felser,bernevig2,ARPES_Huang,thirupathaiah}.

Figure 3(a) displays the electronic band structure of $\gamma-$MoTe$_2$, based on its structure determined at $T = 100$ K, with and without the inclusion of SOC. As previously reported \cite{felser,bernevig2}, electron- and hole-bands intersect along the $\Gamma-X$ direction at energies slightly above $\varepsilon_F$ creating a pair of Weyl type-II points. Figure 3(b) shows a comparison between band structures based on the crystallographic lattices determined at 12 K and at 100 K, respectively. Both sets of electronic bands are nearly identical and display the aforementioned crossings between hole- and electron-bands thus indicating that the electronic structure remains nearly constant below 100 K. Figures 3(c) and 3(d) provide a side perspective and a top view of the overall resulting FS within BZ, respectively. The main features of the DFT calculations are the presence of two-dimensional electron pockets, labeled $e_1$ and $e_2$ in Figs.~3(e) and 3(f) and of large ``star-shaped'' hole-pockets near the $\Gamma-$point, labeled as the $h_2$ and the $h_3$ sheets in Figs.~3(k) and 3(l). These electron and hole pockets nearly ``touch''. Due to the broken inversion symmetry, these bands are not Kramer's degenerate,
and hence the spin-orbit split partners of the corresponding electron and hole pockets are located inside the corresponding bigger sheets.
The $h_1$ hole pocket and the $e_3$ and $e_4$ electron pockets are very sensitive to the position of $\varepsilon_F$ disappearing when $\varepsilon_F$ is moved
by only $\pm 15$ meV.

Figures 4(a) and 4(b) present the angular dependence of the calculated and of the measured FFT spectra of the oscillatory signal (raw data in Fig.~S7 \cite{supplemental}), respectively. In this plot the Onsager relation was used to convert the theoretical FS cross-sectional areas into oscillatory frequencies.
In Fig.~4(b) $\theta$ refers to angles between the $c-$ and the $a-$axis, where $\theta=0^{\circ}$ corresponds to $H\parallel c-$axis,
while $\phi$ corresponds to angles between the $c-$ and the $b-$axis, again relative to the $c-$axis.
As seen, there are striking differences between both data sets with the calculations predicting far more frequencies than the measured ones.
More importantly, for fields oriented from the \emph{c}-axis towards either the $a-$ or the $b-$axis, one observes the complete absence of experimental frequencies around $\sim 1$ kT which, according to the calculations, would correspond to the cross-sectional areas of the hole-pockets $h_2$ and $h_3$. In addition, while many of the predicted electron orbits show a marked two-dimensional character, diverging as the field is oriented towards the $a-$ or the $b-$axis, the experimentally observed frequencies show finite values for fields along either axis. This indicates that these orbits are three-dimensional in character, despite displaying frequencies close to those predicted for the $e_1$ and the $e_2$ pockets for fields along the $c-$axis. These observations, coupled to the non-detection of all of the predicted orbits, in particular the large hole $h_2$ and $h_3$ Fermi surfaces, indicate unambiguously that the actual geometry of the FS of $\gamma-$MoTe$_2$ is different from the calculated one. Notice that frequencies inferior to $F = 100$ T, which correspond to the smaller electron- and hole-pockets and which are particularly sensitive to the position of $\varepsilon_F$ as previously mentioned, were not included in Fig. 4(a) for the sake of clarity.

The calculation shows a significant difference between the SOC-split theoretical bands, which is highlighted by the absence of a frequency around 0.5 kT associated with $e_2$ pocket, along with its presence in association with the $e_1$ pocket. This contrast between both orbits is due to the presence of a ``handle-like'' structure (see Fig. 3(e)) in $e_1$ which gives a maximum cross-section close to the BZ edge. However, at this position there is no maximum cross-section within the BZ for the $e_2$ pocket since its ``handle'' is missing (see Fig.~3(f)). This marked difference in topology between the FSs of both spin-orbit split partners indicates that the strength of the SOC provided by the DFT calculations tends to be considerably larger than the one implied by our experiments. In fact, from the twin peaks observed in the experimental FFT spectra having frequencies around 250 T for fields along the \emph{c}-axis, which are likely to correspond to SOC-split bands due to their similar angular dependence, we can infer that the actual SO-splitting is far less significant than the value predicted by the calculations. We have investigated the possibility of an overestimation of the strength of the SOC within our calculations
which is the mechanism driving the DFT prediction of a large number of dHvA frequencies displaying remarkably different angular dependencies.
For instance, we calculated the angular dependence of the FSs without the inclusion of SOC. This leads to just one, instead of a pair of distinct SOC-split bands, which in fact display angular dependencies very similar to those of orbits $h_3$, $e_1$ and $e_3$ in Fig.~4(a). Notice that part of the discrepancy is attributable to the inter-planar coupling which is not well captured by the DFT calculations \cite{Son}. DFT suggests that this compound is van der Waals like by predicting several two-dimensional (i.e. cylindrical like) FS sheets, when the experiments indicate that the overall FS displays a marked three-dimensional character. This indicates that the inter-planar coupling is stronger than implied by DFT. In any case, from Figs.~4(a) and 4(b) and the above discussion, it is clear that there are significant discrepancies between the calculated and the measured FSs.
Given that the proposed Weyl type-II scenario \cite{bernevig,felser,bernevig2} hinges on a possible touching between electron- and hole-pockets, it is critical to understand their exact geometry, or the reason for the disagreement between predictions and experiments, before one can make any assertion on the existence of the Weyl type-II points in $\gamma-$MoTe$_2$.

To understand the source of the disagreement between calculations and our measurements, we now focus on a detailed comparison between our DFT calculations and a selection of ARPES studies. Figure 5(a) corresponds to data from Ref. \onlinecite{thirupathaiah} depicting an ARPES energy distribution map (EDM) along $k_y$ while keeping $k_x =0$. Figure 5(b) plots its derivative. In both figures the $\Gamma-$point corresponds to $k_y=0$. According to the calculations, this EDM should reveal two valence bands intersecting $\varepsilon_F$ around the $\Gamma-$point; the first leading to two small hole-pockets, or the $h_1$ sheets at either side of $\Gamma$, with the second SOC-split band producing the larger $h_2$ and $h_3$ sheets. Instead, ARPES observes just one band intersecting $\varepsilon_F$ which leads to a single FS sheet of cross-sectional area $S_{\text{FS}} \sim \pi (0.1 \text{ \AA}^{-1})^2$ as indicated by the vertical blue lines in Fig.~5(b). This observation by ARPES questions the existence of the band (or of its intersection with $\varepsilon_F$) responsible for the large $h_2$ and $h_3$ hole-pockets with this band being the one previously reported to touch the electron band that produces the $e_1$ pocket and creating in this way the Weyl type-II points \cite{ARPES_Huang, ARPES_Deng, ARPES_Jiang,ARPES_Tamai,ARPES_Belopolski}. Notice that our dHvA measurements do not reveal any evidence for the original $h_2$ and $h_3$ sheets, thus, being in agreement with this ARPES observation. Furthermore, the Onsager relation $F= S(\hbar/2\pi e)$ yields a frequency of $\sim 330$ T for $S_{\text{FS}}$ which, in contrast, is close to the frequencies observed by us for $\mu_0H \| c-$axis. Figure 5(c) displays the band structure calculated ``ribbons'' obtained by projecting the $k_z$ dependence of the bands onto the $k_x-k_y$ plane. This representation of the band-structure provides a better comparison with the ARPES EDMs.
As seen, there is a good overall agreement between the ARPES and the DFT bands, as previously claimed \cite{ARPES_Huang, ARPES_Deng, ARPES_Jiang,ARPES_Tamai,ARPES_Belopolski}, except for the exact position of $\varepsilon_F$. The purple line depicts the position of $\varepsilon_F$ according to the DFT calculations while the white line depicts the position of $\varepsilon_F$ according to ARPES. A nearly perfect agreement between DFT and ARPES is achievable by shifting the DFT valence bands by $\sim - 50$ meV, which is what ends  suppressing the $h_2$ and $h_3$ FS hole sheets from the measured ARPES EDMs. As shown through Figs.~5(d) and 5(e), this disagreement between the ARPES and the DFT bands is observed in the different ARPES studies\cite{ARPES_Tamai}. Figure 5(d) corresponds to an EDM along the $k_x$ $(k_y=0)$ direction of the BZ. As shown in Fig.~5(e), DFT reproduces this EDM quite well. Nevertheless, as indicated by the yellow lines in both figures, which are positioned at the top of the deepest valence band observed by ARPES, the ARPES bands are displaced by $\sim -45$ meV with respect to the DFT ones. Red dotted lines in Fig.~5(d) indicate the cross-sections of the observed electron pockets or $\sim \pi (0.1 \text{ \AA}^{-1})^2$. Therefore, to match our main dHvA frequencies, the electron bands would have to be independently and slightly displaced towards higher energies to decrease their cross-sectional area. To summarize, DFT and ARPES agree well on the overall dispersion of the bands of $\gamma-$MoTe$_2$, but not on their relative position with respect to $\varepsilon_F$.

Therefore, guided by ARPES, we shifted the overall valence bands of $\gamma-$MoTe$_2$, shown in Figs.~3(a) and 3(b), by -50 meV and the electron ones by +35 meV to recalculate the FS cross-sectional areas as a function of field orientation relative to the main crystallographic axes. The comparison between the measured dHvA cross-sectional areas and those resulting from the shifted DFT bands are shown in Fig.~6. Figure 6(a) displays the Fourier spectra, previously shown in Fig.~4(b), with superimposed colored lines identifying shifted electron (magenta) and hole (blue) orbits according to Fig.~6(b) which displays these frequencies as a function of field orientation for shifted non-SOC-split DFT bands. As seen, the qualitative and quantitative agreement is good, but not perfect. In contrast, Fig.~6(c) displays these orbits/frequencies as a function of field orientation for SOC-split DFT bands. Clearly, and as previously discussed, the approach used to evaluate the effect of the SOC in $\gamma-$MoTe$_2$ seems to overestimate it for reasons that remain to be clarified.
Concerning the Weyl physics in $\gamma-$MoTe$_2$, the displacement of the bands, introduced here to explain our observations based on the guidance provided by previous ARPES studies, would eliminate the crossings between the electron- and the hole-bands as shown in Fig.~7(a). Finally, Figs.~7(b) to 7(g) display the geometry of the Fermi surface resulting from the shifted bands. Overall, the FS displays a distinctly more marked three-dimensional character, with the electrons and the-hole pockets remaining well-separated in \emph{k}-space.
This three-dimensionality is consistent with the observations of Ref. \onlinecite{ARPES_Jiang} which finds that the electronic bands do disperse along the $k_z-$direction implying that $\gamma-$MoTe$_2$ cannot be considered a van der Waals coupled solid. Notice that DFT tends to underestimate the inter-planar coupling in weakly coupled compounds which is at the heart of the disagreement between the calculations and our observations.

\section{Conclusions}
In conclusion, quantum oscillatory phenomena reveal that the geometry of the Fermi surface of $\gamma-$MoTe$_2$ is quite distinct from the one predicted by previous electronic band-structure calculations. Our low-temperature structural analysis \emph{via} synchrotron X-ray diffraction measurements indicates the absence of an additional structural transition below the monoclinic to orthorhombic one that would explain this disagreement, while heat-capacity measurements provide no evidence for an electronic phase-transition upon cooling. In contrast, a direct comparison between DFT calculations and the band-structure reported by angle resolved photoemission spectroscopy reveals a disagreement on the position of the valence bands relative to the Fermi-level, with the experimental valence bands shifted by $\sim -50$ meV relative to the DFT ones. Therefore, one should be careful concerning the claims of a broad agreement between the calculations and the electronic bands revealed by ARPES measurements \cite{ARPES_Huang, ARPES_Deng, ARPES_Jiang, ARPES_Liang, ARPES_Xu, ARPES_Tamai}.

Here, we show that it is possible to describe the angular-dependence of the observed de Haas-van Alphen Fermi surface cross-sectional areas by shifting the position of the DFT bands relative to the Fermi level as indicated by ARPES. However, with this adjustment, the Weyl points, which result from band-crossings that are particularly sensitive to small changes in the lattice constants, are no longer present in the band-structure of $\gamma-$MoTe$_2$. Although our approach of modifying the band structure in order to obtain an agreement with both ARPES and de Haas-van Alphen experiments has only a phenomenological basis, our findings do shed a significant doubt on the existence of the Weyl points in the electronic band structure of $\gamma-$MoTe$_2$. 

Finally, this study combined with the ARPES results in Ref. \onlinecite{daniel2}, indicate that there ought to be a Lifshitz-transition \cite{wu} upon W doping in the $\gamma-$Mo$_{1-x}$W$_x$Te$_2$ series, leading to the disappearance of the central hole pockets in $\gamma-$MoTe$_2$ in favor of the emergence of hole-pockets at either side of the
$\Gamma-$point in $\gamma-$Mo$_{1-x}$W$_x$Te$_2$.

\begin{acknowledgments}
We acknowledge helpful discussions with R. M. Osgood and A. N. Pasupathy.
J.Y.C. is supported by NSF-DMR-1360863.
L.~B. is supported by DOE-BES through award DE-SC0002613
for experiments under high magnetic fields and at very low temperatures,
and by the U.S. Army Research Office MURI Grant W911NF-11-1-0362 for the synthesis
and physical characterization of two-dimensional materials and their heterostructures.
We acknowledge the support of the HLD-HZDR, member of the European Magnetic Field Laboratory (EMFL).
Research conducted at the Cornell High Energy Synchrotron Source (CHESS) is supported by the NSF \& NIH/NIGMS via NSF award DMR-1332208.
The NHMFL is supported by NSF through NSF-DMR-1157490 and the
State of Florida.
\end{acknowledgments}

\end{document}